\documentstyle[12pt]{article}
\title{A new algorithm for automatic photopeak searches}
\author{Z.K.~Silagadze \vspace*{3mm} \\
\small \em
Budker Institute of Nuclear Physics \\
\small \em 630 090, Novosibirsk, Russia  \vspace*{3mm} \\}
\date{}

\begin{document}
\large
\maketitle

\begin{abstract}
A new, "quantum mechanical" algorithm is proposed for automatic
photopeak location in gamma-ray spectra from semiconductor and
scintillator detectors.
\end{abstract}

\section{Introduction}
A great variety of programs can be found in the literature for
automatic gamma-ray spectrum analysis with computers \cite{1}.
An important ingredient of every such program is some peak finding
procedure. Usually the following two characteristics of a local
maximum of continuous function are used in peak searching algorithms:
\newline 1) Near the maximum the curve is convex and so its second
derivative becomes negative, having a minimum value around the maximum
of the peak.
\newline 2) While passing a peak the first derivative changes sign.

The use of the second derivative for automatic photopeak location was
proposed by Mariscotti \cite{2}. The real spectra, however, can have
significant fluctuations because of the discrete and statistical nature
of the data. So a discrete analog of the second derivative (the second
difference) $dd_i=N_{i+1}-2N_i+N_{i-1}$ should be used, $N_i$ being
the pulse heights of the spectrum in channel i.

To reduce further the influence of the statistical fluctuations, this
second difference can be averaged and after n iterations replaced by
the generalized second difference
\begin{eqnarray}
dd_i(n,m)=\sum_{j_n=i-m}^{i+m} \, \sum_{j_{n-1}=j_n-m}^{j_n+m}
\cdots \sum_{j_1=j_2-m}^{j_2+m} \, = \sum \, C_jN_j
\nonumber \end{eqnarray}
\noindent
Its standard statistical deviation is $sd_i(n,m)=\sqrt{\sum C_j^2N_j}$
and therefore not only $dd_i(n,m)<0$ condition is demanded for the peak
area location but also a big enough value of $\left | \frac{dd_i(n,m)}
{sd_i(n,m)} \right | $.

In \cite{3} some modification of this method was used. Instead of
smoothed second difference one can consider the sum $dd_i(k)=
\sum_{j=i-k}^{i+k} \, C_jN_j$ where the $C_j$ coefficients are chosen
in such a way that $dd_i(k)<0$ near the peak. For example in \cite{3}
they take
\begin{eqnarray}
C_j=\frac{j^2-p^2}{p^4}\exp{\left (-\frac{j^2}{2p^2}\right )}=
\frac{d^2}{dx^2} \, \left . \exp {\left (-\frac{x^2}{2p^2}\right )}
\right |_{x=j}
\hspace*{2mm} ,
\nonumber \end{eqnarray}
\noindent
p being some parameter to be optimized. Note that near the peak
\begin{eqnarray} &&
\sum C_jN_j \approx \int_{-\infty}^{\infty} \frac{d^2}{dx^2}
\left [ \exp {\left (-\frac{x^2}{2p^2}\right )} \right ] N(x) \, dx=
\nonumber \\ &&
\int_{-\infty}^{\infty} \exp {\left (-\frac{x^2}{2p^2}\right )}
\frac{d^2N(x)} {dx^2} < 0  \hspace*{2mm} .
\nonumber \end{eqnarray}

In \cite{4} the smoothed first derivative was used for photopeak
searches. The peak is identified at the point where the first
derivative changes sign from positive to negative.

One can smooth not the first or second derivatives but the original
data and establish the sign of the first derivative by simply
comparing consecutive channels. To increase statistical reliability,
one demands not only $N_i>N_{i-1}$, but also $N_{i+1}>N_i$ \cite{5}.
As is claimed in \cite{6}, for low statistics the best result is
given by the following smoothing
\begin{eqnarray}
{\bar N}_i=\frac{1}{9}\left ( N_{i+2}+2N_{i-1}+3N_i+2N_{i+1}+
N_{i+2} \right ) \hspace*{2mm} .
\nonumber \end{eqnarray}

One more method for photopeak position finding, although somewhat
complex, is fitting the whole spectrum or its part by appropriate
analytic function. A lot of various functions were suggested for
this purpose, some of them modeling even Compton continuum \cite{7}.
If resolution is very good, as it is usually for semiconductor detectors,
so that one has almost ideal gaussian peaks and the background in the
peak area is nearly constant, it is possible to fit successive small
sections of the data by Gaussian function and identify the peak position
where the amplitude of the fitted Gaussian becomes significant \cite{8}.

In this note a new and simple enough algorithm is proposed for photopeak
searches, which is completely different from the above described methods.

\section{The idea}
The proposed method is based on a very simple but nice idea, which can
be explained as follows. Suppose we place a small ball on the edge of the
irregular potential wall:

\vspace*{5cm}
\noindent
Classical ball will stop in front of the first obstacle. But if it is
a quantum one and can penetrate through narrow barriers it still goes down
to the potential wall bottom and oscillates there.

For automatic photopeak finding this idea can be realized, for example,
in such a way. Let us take a point somewhere on the right slope of the
peak and let this point can jump left or right by one channel, or remain
on its position, so that the probability to jump left is proportional
to $\exp {\left [ \frac{N_{i-1}-N_i}{\sqrt{N_{i-1}+N_i}} \right ]}$ and
to jump right \-\-\- $\exp {\left [ \frac{N_{i+1}-N_i}{\sqrt{N_{i+1}+N_i}}
\right ]}$. This point will climb up quickly enough to the photopeak and
will oscillate on the top. If now we follow up some amount of jumps and
calculate the mean position, this just gives us the photopeak location.

To smear the "quantum ball" even more, one can take for the probability
to jump from channel i to channel i-1 :
\begin{eqnarray}
P_{i,i-1} \sim \sum_{k=1}^{m} \exp { \left [ A_k \frac{N_{i-k}-N_i}
{\sqrt{N_{i-k}+N_i}} \right ] } \hspace*{2mm} ,
\nonumber \end{eqnarray}
\noindent
where $A_k$ are some numerical constants. Changing them and the number
m, one can govern the penetrating ability of the ball.

\section{Realization via discrete Markov chain}
In fact the above described set of channels and transition probabilities
can be treated as a finite Markov chain \cite{9,10}.

For simplicity let us assume that from any state (channel) in this Markov
chain only the closest left and right neighbour states can be reached
with nonzero probabilities in one step, and  that the probability for
any state to remain unchanged is zero. Thus the transition probability
matrix for this chain looks like
\begin{eqnarray}
P=\left ( \begin{array}{cccccccc}
0 & 1 & 0 & 0 & 0 & \cdot & \cdot & \cdot \\
P_{21} & 0 & P_{23} & 0 & 0 & \cdot & \cdot & \cdot \\
0 & P_{32} & 0 & P_{34} & 0 & \cdot & \cdot & \cdot \\
\cdot & \cdot & \cdot & \cdot & \cdot & \cdot & \cdot & \cdot \\
0 & \cdot & \cdot & \cdot & \cdot & 0 & 1 & 0
\end{array} \right ) \hspace*{2mm} .
\label{eq1} \end{eqnarray}
\noindent
Where we shall take
\begin{eqnarray}
P_{i,i \pm 1} = A_i \, \sum_{k=1}^{m} \exp { \left [ \frac{N_{i \pm k}
-N_i}{\sqrt{N_{i \pm k}+N_i}} \right ] } \hspace*{2mm} ,
\label{eq2} \end{eqnarray}
\noindent
$A_i$ normalization constant is defined from the $P_{i,i-1}+P_{i,i+1}=1$
condition.

This Markov chain has a very simple invariant distribution \cite{10}:
\begin{eqnarray}
u_2=\frac{P_{12}}{P_{21}}u_1 \; , \; u_3=\frac{P_{12}P_{23}}{P_{32}
P_{21}}u_1 \; , \; \cdots \; , \; u_n=\frac{P_{12}P_{23} \cdots
P_{n-1,n}}{P_{n,n-1}P_{n-1,n-2} \cdots P_{21}}u_1 \hspace*{2mm} ,
\label{eq3} \end{eqnarray}
\noindent
$u_1$ being defined from the normalization condition
\begin{eqnarray}
\sum_{i=1}^n u_i = 1 \hspace*{3mm} .
\label{eq4} \end{eqnarray}

Now this invariant distribution has very sharp peaks which correspond
to a local maximums in the original spectrum. This is illustrated by
Fig.1. It is much simpler to identify peaks by some computer program in
the $u_n$ distribution than in the original data. Fig.2 gives a clear
example of this (note the logarithmic scale on the vertical axis).

In fact this method of photopeak finding can work even for very low
statistics, as Fig.3 shows.

\section{Photopeak fitting}
After the photopeak is located by the above described algorithm, one
can fit it by gaussian function and extract its other characteristics.
Fig.2 and Fig.3 show that our "quantum"algorithm not only gives the
peak position, but can also provide useful limits in which it is
worthwhile  to fit it by Gaussian.

For good enough gaussian peaks one can use a non-iterative method
for the fast fitting \cite{11}. Suppose we want to fit a Gaussian
to the spectrum region from channel $m_1$ to channel $m_2$. Let us
form an array $S(i)=\ln{\frac{N_i}{N_{i+m}}}-\ln{\frac{N_{i-1}}
{N_{i+m-1}}}$, where m is some integer number. If we have a pure
Gaussian
\begin{eqnarray}
N_i=A \exp{\left [ -\frac{(i-p)^2}{2\sigma^2} \right ] }
\hspace*{2mm} ,
\nonumber \end{eqnarray}
\noindent
then $S(i)=\frac{m}{\sigma^2}$ doesn't depend on i. In fact this
will not be the case because of statistical fluctuations and we
should find a constant C which minimizes the sum
\begin{eqnarray}
\sum_{i=m_1+1}^{m_2-m} \frac{[S(i)-C]^2}{D^2(i)}
\; \; , \; \; D^2(i)=\frac{1}{N_i}+\frac{1}{N_{i+m}}+
\frac{1}{N_{i-1}}+\frac{1}{N_{i+m-1}} \; \; ,
\nonumber \end{eqnarray}
\noindent
where $D(i)$ is a standard deviation for the casual quantity $S(i)$.
The solution of this linear minimization problem is well known
\begin{eqnarray}
C=\left [ \sum_{i=m_1+1}^{m_2-m} \frac{S(i)}{D^2(i)} \right ]
\bigg/ \left [ \sum_{i=m_1+1}^{m_2-m} \frac{1}{D^2(i)} \right ]
\; \; . \label{eq5} \end{eqnarray}
\noindent
If we disregard statistical fluctuations of $D(i)$, the statistical
error in determination of C is
\begin{eqnarray}
\Delta C= \left [ \sum_{i=m_1+1}^{m_2-m} \frac{1}{D^2(i)} \right ]
^{-1/2} \; \; .
\nonumber \end{eqnarray}
\noindent
So we get
\begin{eqnarray}
\sigma=\sqrt{\frac{m}{C}} \; \; , \; \; \Delta \sigma=
\frac{\sigma^3}{2m} \Delta C
\label{eq6} \end{eqnarray}
The parameter m can be determined from the condition $\Delta \sigma$
to be minimal. Let us take as an estimation $D(i)=\frac{2}{\sqrt{N_i}
}$. Then $ \Delta \sigma \sim \frac{1}{m\sqrt{m_2-m_1-m}}$, which is
minimal for
\begin{eqnarray}
m=\frac{2}{3}(m_2-m_1) \; \; .
\label{eq7} \end{eqnarray}
Analogously one can find the position and amplitude of the fitted peak
by considering arrays $P(i)=\ln{\frac{N_i}{N_{i+m}}}=\frac{m(2i-2p+m)}
{2\sigma^2}$ and $Q(i)=\ln{N_i}=\ln{A}-\frac{(i-p)^2}{2\sigma^2}$.
The results are
\begin{eqnarray} &&
p=\left [ \sum_{i=m_1+1}^{m_2-m} \frac{i-\frac{\sigma^2}{m}P(i)
+\frac{m}{2}}{DP(i)^2} \right ]
\bigg/ \left [ \sum_{i=m_1+1}^{m_2-m} \frac{1}{DP(i)^2} \right ]
\nonumber \\ &&
DP(i)^2=\frac{\sigma^2}{m^2} \left [4(\Delta \sigma)^2P^2(i)+
\sigma^2 \left ( \frac{1}{N_i}+\frac{1}{N_{i+m}} \right ) \right ]
\; \; ,
\label{eq8} \end{eqnarray}
\noindent
and
\begin{eqnarray} &&
\ln{A}=\left [ \sum_{i=m_1}^{m_2} \frac{\ln{N_i}+\frac{(i-p)^2}{2\sigma^2}}
{DN(i)^2} \right ]
\bigg/ \left [ \sum_{i=m_1}^{m_2} \frac{1}{DN(i)^2} \right ]
\nonumber \\ &&
DN(i)^2=\frac{1}{N_i}+\left( \frac{\Delta \sigma}{\sigma} \right )^2
\left ( \frac{i-p}{\sigma} \right )^4
\; \; .
\label{eq9} \end{eqnarray}
\noindent
where m is defined by eq.7 and $\frac{\Delta p}{p} \ll
\frac{\Delta \sigma}{\sigma}$ relation was assumed while deriving
the last equation.

The degree of fit between the calculated and observed spectra can be
further analyzed by means of some quantitative criteria, suggested in
the literature \cite{12}.

\section{Conclusions}
It seems to us that the proposed algorithm is simple and effective
enough to be recommended for practical applications.

An earlier version of it was used with excellent results for uniformity
studies of about 1650 NaJ crystals for SND detector \cite{13}, and
also for vacuum phototriodes testing \cite{14}.

\section*{Acknowledgements}
The author is grateful to V.B.~Golubev, S.I.~Serednyakov and
A.N.~Perysh\-kin for useful discussions.

\newpage
\section*{Figure captures}

\noindent
Fig.1 --- Invariant probability distribution u(i) for single
gaussian peak, m=3.

\vspace*{3mm}
\noindent
Fig.2 --- Invariant probability distribution for peak + background,
m=3.

\vspace*{3mm}
\noindent
Fig.3 --- The same as on Fig.2 for low statistics.

\end{document}